# The Scale of Universalization in Multiplicative Kantian Equilibrium

Igor Sloev[1 2], Gerasimos Lianos[3]

Multiplicative Kantian equilibrium requires agents to evaluate whether they would prefer all players to change their actions by a common proportional factor. This note argues that such a counterfactual is not fully specified by the material payoff environment alone: one must also specify the scale in which proportional universalization is performed. Building on the Sher–Roemer discussion of strategic non-invariance, we characterize how a smooth monotone reparametrization of actions changes the multiplicative Kantian first-order condition. We then illustrate the issue in a simple two-country abatement problem. Universalizing physical abatement and universalizing abatement expenditures are both natural interpretations of "doing one's part," but they lead to different Kantian outcomes, even though they describe the same material environment. The note also identifies special cases in which the interior first-order condition is representation-invariant, including common-exponent power transformations and symmetric profiles under identical transformations. The conclusion is constructive: applying multiplicative Kantian equilibrium to an economic environment requires specifying both the use of the Kantian criterion and the represented scale in which the Kantian counterfactual is evaluated.

[1]School of Computational Social Sciences, European University at St. Petersburg, Russia, ORCID: 0000-0003-0286-1034, sia@eu.spb.ru.

[2]Corresponding author: 6/1A Gagarinskaya st., St. Petersburg 191187 Russia, sia@eu.spb.ru.
[3]Higher Colleges of Technology, Abu Dhabi College, Baniyas, Abu Dhabi, UAE, glianos@hct.ac.ae

## 1. Introduction

Multiplicative Kantian equilibrium has become an important way of modeling cooperative conduct without abandoning self-interested preferences. In Roemer's formulation, a strategy profile is Kantian if no agent would prefer that all agents change their contributions by the same multiplicative factor (Roemer 2010, 2015, 2019). In this sense, Kantian optimization replaces the unilateral-deviation logic of Nash equilibrium with a universalized counterfactual: an agent asks what she would prefer if everyone changed action in the same way. In a variety of environments with externalities, this shift yields more cooperative and often more efficient outcomes than Nash behavior.

A difficulty, however, was highlighted by Sher (2020). Kantian equilibrium is not strategically invariant in the way Nash equilibrium is: payoff-preserving relabelings of strategy spaces may leave Nash predictions unchanged while altering Kantian ones. Sher argues that this means Kantian equilibrium requires more than preference data; it requires a normatively privileged way of measuring strategies. Roemer (2020) accepts the non-equivalence point, but interprets it differently. For him, the dependence on representation is not merely a defect: Kantian cooperation requires identifying the relevant symmetry in the situation, and the way strategies are described is therefore part of the moral content of the model.

This note starts from that Sher–Roemer agreement. Its aim is not to rediscover strategic non-equivalence. Rather, the aim is to show that the issue has substantive bite in a very ordinary collective-action setting. In many economic environments, there is not a single obvious scale in which proportional universalization should be defined. In a mitigation problem, for example, agents may universalize physical abatement, monetary abatement cost, or some other burden-sharing variable. The relevance of such alternative measures is familiar from climate-policy discussions of effort sharing and fair burden allocation (Ringius, Torvanger, and Underdal 2002; Höhne, den Elzen, and Escalante 2014; Böhringer and Helm 2008). These descriptions can be materially equivalent in the sense that they generate the same feasible allocations and payoff consequences, yet they encode different answers to the Kantian question: what would be desirable if all of us did the same proportionately?

The claim is narrower than a general critique and stronger than a remark about notation. We do not argue that agents strategically manipulate the scale in which the Kantian criterion is applied. Nor do we argue that Kantian equilibrium is ill-defined once a represented game is fixed. The point is instead that a material collective-action environment need not determine a unique Kantian representation. For Nash equilibrium, one-to-one monotone reparametrizations of individual choice variables are innocuous because unilateral deviations are preserved. For multiplicative Kantian equilibrium, they need not be innocuous, because the solution concept evaluates coordinated proportional deviations, and proportionality is representation-dependent. Thus, the scale of universalization is not merely notation; it is part of the specification of the Kantian decision problem.

We develop this idea in a two-country abatement model with the standard public-good structure familiar from the literature on international environmental agreements: countries bear their own abatement costs, while benefits depend on aggregate abatement (Barrett 1994, 2003). The model is used only as a transparent collective-action environment. We compare two common scales of universalization. Under an abatement representation, countries ask whether they would prefer that all countries scale up their physical emission reductions proportionately. Under a cost representation, they ask whether they would prefer that all countries scale up their abatement spending proportionately. Both representations are shared by all players; neither relies on heterogeneous moral languages across agents. Yet they may lead to different multiplicative Kantian outcomes. The contrast is therefore not between a common Kantian criterion and manipulation, but between two plausible interpretations of what it means for all parties to "do their part."

The note makes two contributions. First, it characterizes how a smooth monotone change of variables alters the multiplicative Kantian first-order condition. This identifies the mechanism through which representation affects the equilibrium and clarifies why some transformations preserve the multiplicative Kantian first-order condition while others do not. Second, it shows that the issue raised by Sher and Roemer is not confined to abstract relabelings. It appears in an applied environment where alternative scales are economically natural and normatively interpretable: physical contribution and monetary sacrifice express different candidate symmetries of collective action.

This perspective suggests a modest but important methodological lesson. Applications of multiplicative Kantian equilibrium should distinguish the underlying collective-action environment from the represented Kantian game to which the equilibrium concept is applied. The former specifies feasible physical choices, payoff consequences, and technological relations among variables; the latter must also specify the scale in which proportional universalization is evaluated. In that respect, the relevant comparison is not simply Nash versus Kantian equilibrium, but Nash versus Kantian equilibrium under a stated representation of collective action. Making this modeling choice explicit does not weaken the Kantian program. On the contrary, it clarifies what is required to apply it in environments where several natural symmetries coexist.

The rest of the note is brief. Section 2 formalizes the role of the scale of universalization and states the transformation result for multiplicative Kantian equilibrium. Section 3 illustrates the point in the two-country abatement example. Section 4 identifies special cases in which the interior first-order condition is representation-invariant. Section 5 concludes.

## 2. The scale of universalization

Before turning to the example, we establish a general proposition relating multiplicative Kantian first-order conditions under alternative scales of universalization.

Let $I = 1, \ldots, n$. Each player $i$ has an action variable $q_i$ in an interval $Q_i \subset R_+$. We focus on interior profiles $q^*$ with $q_i^* > 0$ for all $i$. Let $q = (q_1, \ldots, q_n)$, and let $u_i(q)$ be continuously differentiable on the relevant domain. In the multiplicative Kantian formulation, player $i$ evaluates the one-dimensional path along which all players multiply their actions by the same factor $\lambda$: $u_i(\lambda q_1, \ldots, \lambda q_n)$.

An interior profile $q^*$ is a multiplicative Kantian equilibrium, in this representation, if $\lambda = 1$ maximizes $u_i(\lambda q_1^*, \dots, \lambda q_n^*)$ over all λ such that $(\lambda q_1^*, \dots, \lambda q_n^*) \in Q_1 \times \cdots \times Q_n$, for every player $i$.

This condition makes explicit that the multiplicative Kantian criterion uses the representation in which the expression "multiply all actions by the same factor" is defined.

Now consider a smooth monotone one-to-one reparametrization of each player's action, $z_i = f_i(q_i)$, where $f_i$ is continuously differentiable, strictly increasing, invertible, and satisfies $f_i(0) = 0$. Let $q_i = f_i^{-1}(z_i)$ and define the transformed payoff function by

$$v_i(z_1, \dots, z_n) = u_i\big(f_1^{-1}(z_1), \dots, f_n^{-1}(z_n)\big).$$

The underlying environment is unchanged: every profile $z$ corresponds to exactly one profile $q$, and material payoffs coincide. However, multiplicative Kantian deviations in the z-representation are different. They are given by $z \rightarrow \lambda z$, or, in the original q-coordinates, $q_i \rightarrow f_i^{-1}\big(\lambda f_i(q_i)\big)$.

The following proposition records the induced first-order condition.

**Proposition.** Let $q^*$ be an interior profile and let $z^* = f(q^*)$. In the q-representation, the interior multiplicative Kantian first-order condition for player $i$ is

$$\sum_{j=1}^{n} \big(\partial u_i(q^*)/\partial q_j\big) q_j^* = 0.$$

In the z-representation, the corresponding first-order condition is

$$\sum_{j=1}^{n} \big(\partial u_i(q^*)/\partial q_j\big)\big[f_j\big(q_j^*\big)/f_j'\big(q_j^*\big)\big] = 0.$$

**Proof.**

In the q-representation, define $\phi_i(\lambda) = u_i(\lambda q_1^*, \dots, \lambda q_n^*)$. By the chain rule,

$$\phi_i'(1) = \sum_{j=1}^{n} \big(\partial u_i(q^*)/\partial q_j\big) q_j^*.$$

Thus the first-order condition $\phi_i'(1) = 0$ gives $\sum_{j=1}^{n} \big(\partial u_i(q^*)/\partial q_j\big) q_j^* = 0$.

In the z-representation, define $\psi_i(\lambda) = v_i(\lambda z_1^*, \dots, \lambda z_n^*)$. Since $v_i(z) = u_i\big(f_1^{-1}(z_1), \dots, f_n^{-1}(z_n)\big)$, we have $\psi_i(\lambda) = u_i\big(f_1^{-1}(\lambda z_1^*), \dots, f_n^{-1}(\lambda z_n^*)\big)$.

Applying the chain rule again,

$$\psi_i'(1) = \sum_{j=1}^{n} \big(\partial u_i(q^*)/\partial q_j\big) d/d\lambda\, f_j^{-1}\big(\lambda z_j^*\big)|_{\lambda=1}.$$

Since $z_j^* = f_j\big(q_j^*\big)$,

$$d/d\lambda f_j^{-1}(\lambda z_j^*)|_{\lambda=1} = (f_j^{-1})'(z_j^*)z_j^* = z_j^*/f_j'(q_j^*) = f_j(q_j^*)/f_j'(q_j^*).$$

Hence $\psi_i'(1) = \sum_{j=1}^{n}(\partial u_i(q^*)/\partial q_j)[f_j(q_j^*)/f_j'(q_j^*)]$. The transformed first-order condition $\psi_i'(1) = 0$ is $\sum_{j=1}^{n}(\partial u_i(q^*)/\partial q_j)[f_j(q_j^*)/f_j'(q_j^*)] = 0$.

This proves the claim.

The proposition characterizes how the multiplicative Kantian first-order condition changes under a smooth monotone reparametrization. The transformation affects the condition through the weights $f_j(q_j^*)/f_j'(q_j^*)$, which replace the original weights $q_j^*$. In the example below, strict concavity in lambda ensures that these first-order conditions are also sufficient for an MKE. Section 4 then uses the same proposition to identify special cases in which the interior first-order condition is representation-invariant.

## 3. An abatement example

We now apply the preceding argument to a two-country abatement environment. The purpose of the example is not to model climate policy, but to show that two natural representations of the same underlying abatement environment can generate different multiplicative Kantian equilibria.

There are two countries, $i = 1,2$. Country $i$ can achieve physical abatement $q_i \geq 0$ by incurring total abatement expenditure $C_i \geq 0$. The technological relation between expenditure and physical abatement is $C_i = k_i(exp(q_i) - 1)$, or equivalently $q_i = ln(1 + C_i/k_i)$. We assume $k_1 = 1/5, k_2 = 1/2$. Thus country 1 is the lower-cost abater. Aggregate abatement is $Q = q_1 + q_2$. Country i's payoff is aggregate abatement minus its own abatement expenditure: $U_i = Q - C_i$.

At this stage, the underlying abatement environment has been described: feasible physical choices, abatement expenditures, and the payoffs they generate. What has not yet been fixed is the scale in which the Kantian counterfactual is evaluated. The same environment can be represented in at least two natural ways. In representation A, countries choose physical abatement levels $q_i$. In representation B, countries choose abatement expenditures $C_i$. Since $C_i = k_i(exp(q_i) - 1)$ is strictly increasing, the two representations describe the same feasible physical choices and the same payoff consequences. They differ in the scale of universalization.

### 3.1 Nash benchmark

Before considering Kantian behavior, consider the Nash benchmark. In either representation, country $i$ chooses its own action taking the other country's action as fixed. In the abatement representation, this means solving

$$max_{q_i \geq 0} q_1 + q_2 - k_i(exp(q_i) - 1).$$

The first-order condition is $1 - k_i exp(q_i) = 0$. Therefore $q_i^N = ln(1/k_i)$. For $k_1 = 1/5$ and $k_2 = 1/2$, we obtain $q^N = (ln5, ln2) \approx (1.609, 0.693)$.

The corresponding expenditures are $C_1^N = 0.8, C_2^N = 0.5$. Aggregate abatement is $Q^N \approx 2.302$, and payoffs are $U_1^N \approx 1.503$, $U_2^N \approx 1.803$.

The same material outcome is obtained if the Nash problem is written directly in expenditure coordinates. This is because the two descriptions are linked by a one-to-one transformation of individual choices.

### 3.2 Representation A: universalizing physical abatement

In representation A, the strategy of country $i$ is physical abatement $q_i$. The payoff function is

$$U_i^A(q_1, q_2) = q_1 + q_2 - k_i(exp(q_i) - 1).$$

The multiplicative Kantian counterfactual asks whether country $i$ would prefer all countries to multiply their physical abatement quantities by the same factor $\lambda$. Thus country $i$ evaluates

$$U_i^A(\lambda q_1, \lambda q_2) = \lambda(q_1 + q_2) - k_i(exp(\lambda q_i) - 1).$$

The second derivative with respect to $\lambda$ is

$$d^2 U_i^A(\lambda q_1, \lambda q_2)/d\lambda^2 = -k_i exp(\lambda q_i) q_i^2 < 0.$$

Hence the objective is strictly concave in \lambda, and the first-order condition at $\lambda = 1$ is sufficient. The MKE conditions in representation A are

$$q_1 + q_2 - k_i exp(q_i) q_i = 0, i = 1,2.$$

Equivalently, $Q = k_1 exp(q_1) q_1 = k_2 exp(q_2) q_2$. For $k_1 = 1/5$ and $k_2 = 1/2$, the solution is $q^A \approx (2.154, 1.560)$.

The corresponding expenditures are $C_1^A \approx 1.524, C_2^A \approx 1.881$. Thus $Q^A \approx 3.714$, and $U_1^A \approx 2.190, U_2^A \approx 1.834$.

Representation A therefore captures proportional universalization of physical abatement. In this representation, “doing the same” means changing physical abatement quantities in the same proportion.

### 3.3 Representation B: universalizing abatement expenditure

In representation B, the strategy of country $i$ is abatement expenditure $C_i$. Physical abatement is then determined by $q_i = ln(1 + C_i/k_i)$.

The payoff function in expenditure coordinates is

$$U_i^B(C_1, C_2) = ln(1 + C_1/k_1) + ln(1 + C_2/k_2) - C_i.$$

The multiplicative Kantian counterfactual now asks whether country $i$ would prefer all countries to multiply their abatement expenditures by the same factor $\lambda$. Thus country $i$ evaluates

$$U_i^B(\lambda C_1, \lambda C_2) = ln(1 + \lambda C_1/k_1) + ln(1 + \lambda C_2/k_2) - \lambda C_i.$$

The second derivative with respect to $\lambda$ is

$$d^2U_i^B(\lambda C_1, \lambda C_2)/d\lambda^2 = -C_1^2/(k_1 + \lambda C_1)^2 - C_2^2/(k_2 + \lambda C_2)^2 < 0.$$

Hence the objective is strictly concave in \lambda, and the first-order condition at $\lambda = 1$ is sufficient. The MKE conditions in representation B are

$$C_1/(k_1 + C_1) + C_2/(k_2 + C_2) - C_i = 0, i = 1,2.$$

Since the first two terms are the same in both equations, these conditions imply

$$C_1^B = C_2^B = C^*.$$

Substituting this into the first-order condition gives $1 = 1/(k_1 + C^*) + 1/(k_2 + C^*)$. For $k_1 = 1/5$ and $k_2 = 1/2$, the positive solution is $C^* \approx 1.661$.

Thus $C_1^B = C_2^B \approx 1.661$, $q^B = \left(ln(1 + C^*/k_1), ln(1 + C^*/k_2)\right) \approx (2.231,1.464)$. Therefore $Q^B \approx 3.695$, and payoffs are $U_1^B = U_2^B \approx 2.033$.

In this representation, "doing the same" means changing abatement expenditures in the same proportion.

### 3.4 Comparison

The three outcomes are summarized in Table 1.

Table 1. Nash and Kantian outcomes in the abatement example

| | $q_1$ | $q_2$ | $C_1$ | $C_2$ | $Q$ | $U_1$ | $U_2$ |
|---|---|---|---|---|---|---|---|
| Nash | 1.609 | 0.693 | 0.8 | 0.5 | 2.302 | 1.503 | 1.803 |
| MKE-A | 2.154 | 1.560 | 1.524 | 1.881 | 3.714 | 2.190 | 1.834 |
| MKE-B | 2.231 | 1.464 | 1.661 | 1.661 | 3.695 | 2.033 | 2.033 |

Both Kantian representations improve on Nash in terms of aggregate abatement. Yet they do not coincide. Representation A asks countries to universalize physical contributions. Representation B asks countries to universalize abatement expenditures. Both scales are common to all players, and neither scale is chosen strategically by the players. They are two natural ways of completing the same underlying abatement environment into a represented Kantian game.

The difference between A and B is therefore not a difference in physical possibilities or payoff consequences. It is a difference in the scale in which the Kantian proportionality requirement is imposed. The example illustrates the main point of the paper: multiplicative Kantian equilibrium is well defined once the represented game is fixed, but the underlying collective-action environment does not by itself determine which proportional universalization is relevant.

## 4. Special cases

The example above shows that representation dependence can matter in a simple collective-action problem. The proposition also clarifies why this dependence is neither universal nor arbitrary. The following two observations identify cases in which the change of representation does not alter the relevant multiplicative Kantian first-order condition.

**4.1 Power transformations**

Suppose that $z_i = f_i(q_i) = b_i q_i^r$, where $b_i > 0$ and $r > 0$ is the same for all players. Then $f_i'(q_i) = b_i r q_i^{r-1}$. Therefore $f_i(q_i)/f_i'(q_i) = b_i q_i^r/(b_i r q_i^{r-1}) = q_i/r$.

By the proposition, the transformed first-order condition becomes

$$\sum_{j=1}^{n} \left(\partial u_i(q^*)/\partial q_j\right)\left(q_j^*/r\right) = 0.$$

Since $r > 0$, this is equivalent to

$$\sum_{j=1}^{n} \left(\partial u_i(q^*)/\partial q_j\right) q_j^* = 0.$$

Thus common-exponent power transformations preserve the multiplicative Kantian first-order condition. Intuitively, they preserve multiplicative rays. If $q_i$ is replaced by $b_i q_i^r$, then multiplying $q_i$ by $\lambda$ corresponds to multiplying $z_i$ by $\lambda^r$. The path is the same; only its parameterization changes.

This explains why quadratic costs do not generate the scale effect when the expenditure variable is $z_i = c_i(q_i) = b_i q_i^2$. The move from $q_i$ to $z_i$ is a power transformation, and multiplicative universalization in $z$ is equivalent to multiplicative universalization in $q$.

**4.2 Symmetric profiles under identical transformations**

A second special case arises at symmetric profiles under identical transformations. Suppose the game is symmetric and consider a symmetric interior profile $q^* = (x, \dots, x)$, with $x > 0$. Suppose also that the same reparametrization is applied to each player, $z_i = f(q_i)$.

By the proposition, the transformed first-order condition uses the weights $f\left(q_j^*\right)/f'\left(q_j^*\right)$. At the symmetric profile $q^* = (x, \dots, x)$, these weights are identical across players: $f\left(q_j^*\right)/f'\left(q_j^*\right) = f(x)/f'(x), j = 1, \dots, n$.

Hence the transformed first-order condition is equivalent to

$$\sum_{j=1}^{n} \left(\partial u_i(q^*)/\partial q_j\right) = 0,$$

because $f(x)/f'(x) > 0$. The original first-order condition is also equivalent to the same expression, since it is $\sum_{j=1}^{n} \left(\partial u_i(q^*)/\partial q_j\right) x = 0$ and $x > 0$.

Thus, at symmetric interior profiles under identical transformations, the original and transformed multiplicative Kantian first-order conditions coincide.

This does not mean that representation never matters in symmetric games. Rather, it explains why the issue can be hidden in symmetric examples. If all players are identical, the same transformation is used for everyone, and attention is restricted to the symmetric diagonal, the scale of universalization may not affect the symmetric interior MKE condition. Representation dependence becomes visible when transformations change the relative weights of coordinates along the universalization path, as in asymmetric collective-action environments.

## 5. Concluding remark

The preceding analysis clarifies the role of representation in multiplicative Kantian equilibrium. It does not show that MKE is ill-defined. For any fixed represented game, MKE is well-defined. The point is different: a material collective-action environment does not by itself determine which proportional universalization is relevant.

For Nash equilibrium, payoff-preserving one-to-one reparametrizations of individual choices are behaviorally neutral. The same material situation, described in different but equivalent coordinates, gives the same Nash outcome. For multiplicative Kantian equilibrium, such reparametrizations are neutral only when they preserve the structure of universalized proportional deviations. Power transformations and symmetric profiles under identical transformations are two important cases in which the interior first-order condition is representation-invariant. Outside such cases, the scale of universalization is a substantive part of the model.

Thus the relevant comparison is not simply Nash versus MKE. It is Nash versus MKE under a specified scale of universalization. In applications, one must say not only that agents use the multiplicative Kantian criterion, but also in which variable the Kantian counterfactual "what if all of us did the same?" is evaluated.

**6. Statements and Declarations.** The authors have no relevant financial or non-financial interests to disclose.